\documentstyle[epsfig]{article}

\begin{document}

\def\qed{\relax\ifmmode\hskip2em \Box\else\unskip\nobreak\hskip1em $\Box$\fi}
\newtheorem{lem}{Lemma}
\newtheorem{thm}{Theorem}
\newtheorem{defn}{Definition}
\newenvironment{pf}{\noindent{\bf Proof.}}{}


\newcommand{\congA}{\mathrel{\cong_{\cal A}}}
\newcommand{\congB}{\mathrel{\cong_{\cal B}}}
\newcommand{\congR}{\mathrel{\cong_{\cal R}}}

\renewcommand{\div}{\mathop{\mathrm{div}}}

\newcommand{\Rea}{${\cal R}_{\mathrm{ea}}$}
\newcommand{\Cea}{${\cal C}_{\mathrm{ea}}$}
\newcommand{\RL} {${\cal R}_{\mathrm{pcsp}}$}
\newcommand{\CL} {${\cal C}_{\mathrm{pcsp}}$}

\newcommand\US{\mbox{US}}
\newcommand{\ZN}{${\cal Z}_N$}
\newcommand{\Ztwo}{${\cal Z}_2$}

\def\dx{d\kern -0.32em\raise 0.70ex \hbox{-}\kern-0.051ex \hbox{}}
\def\DX{D\kern -0.7em \raise 0.26ex \hbox{-}\kern 0.829ex \hbox{}}

\newcommand{\If}{{\bf if\ }}
\newcommand{\Then}{{\bf then\ }}
\newcommand{\Else}{{\bf else\ }}
\newcommand{\Block}{{\bf block}}
\newcommand{\EndBlock}{{\bf endblock}}
\newcommand{\Endif}{{\bf endif}}
\newcommand{\Var}{{\bf var\ }}
\newcommand{\Endvar}{{\bf endvar}}
\newcommand{\RangesOver}{{\bf ranges over\ }}
\newcommand{\And}{{\bf and\ }}
\newcommand{\Or}{{\bf or\ }}
\newcommand{\Not}{{\bf not\ }}
\newcommand{\Choose}{{\bf choose\ }}
\newcommand{\In}{{\bf in\ }}
\newcommand{\Endchoose}{{\bf endchoose}}


\title{Equivalence Is In The Eye Of The Beholder%
\thanks{\emph{Theoretical Computer Science}, vol. 179, June 1997, to
appear.}}
\author{Yuri Gurevich%
\thanks{Partially supported
by ONR grant N00014-94-1-1182 and NSF grant CCR-95-04375.  The first
author was with the Centre National de la Recherche
Scientifique, Paris, France, during the final stage of this work.}
and James K. Huggins\footnotemark[2]\\
EECS Department, University of Michigan, Ann
Arbor, MI, 48109-2122, USA.}
\date{}
\maketitle

\begin{abstract}
     In a recent provocative paper, Lamport points out "the 
        insubstantiality of processes" by proving the equivalence of
        two different decompositions of the same intuitive algorithm
        by means of temporal formulas.  We point out that the correct
        equivalence of algorithms is itself in the eye of the beholder.
        We discuss a number of related issues and, in particular, 
        whether algorithms can be proved equivalent directly.
\end{abstract}


\section{Introduction}

This is a reaction to Leslie Lamport's
``Processes are in the Eye of the Beholder'' \cite{lamport}.
Lamport writes:

\begin{quote} \begin{em}
A concurrent algorithm is traditionally represented as the composition
of processes. We show by an example that processes are an artifact
of how an algorithm is represented. The difference between a two-process
representation and a four-process representation of the same algorithm is no
more fundamental than the difference between $2+2$ and $1+1+1+1$.
\end{em} \end{quote}

To demonstrate his thesis, Lamport uses two different programs for a
first-in, first-out ring buffer of size $N$.  He represents the
two algorithms by temporal formulas and proves the equivalence of the
two temporal formulas.  

We analyze in what sense the two algorithms are and are not
equivalent.  There is no one notion of equivalence appropriate for all
purposes and thus the ``insubstantiality of processes'' may itself be
in the eye of the beholder.  There are other issues where we disagree
with Lamport.  In particular, we give a direct equivalence proof for
two programs without representing them by means of temporal formulas.

This paper is self-contained.  In the remainder of this section, we
explain the two ring buffer algorithms and discuss our disagreements
with Lamport.  In Section~\ref{eaintro}, we give a brief
introduction to evolving algebras.  In Section~\ref{ringeas},
we present our formalizations of the ring buffer algorithms as evolving
algebras.  In Section~\ref{equivpf}, we define a version of
lock-step equivalence and prove that our formalizations of these
algorithms are equivalent in
that sense.  Finally, we discuss the inequivalence of these algorithms
in Section~\ref{inequiv}.

\subsection{Ring Buffer Algorithms}

\begin{figure}[htbp]
\centerline{\framebox{\epsfig{file=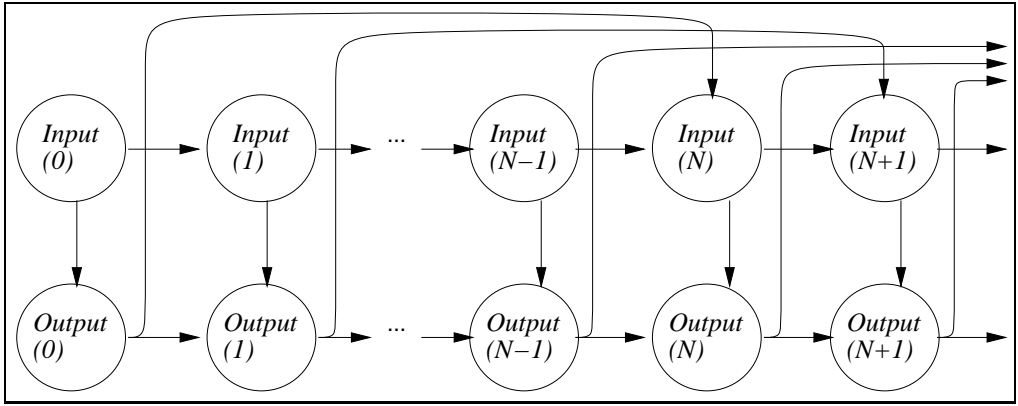,width=.8\textwidth}}}
\caption{Moves of the ring-buffer algorithm.}\label{equivsketch1}
\end{figure}

The ring buffer in question is implemented by means of an array of $N$
elements.  The $i$th input (starting with $i=0$) is stored in slot $i
\bmod N$ until it is sent out as the $i$th output.  Items may be
placed in the buffer if and only if the buffer is not full; of course,
items may be sent from the buffer if and only if the buffer is not
empty.  Input number $i$ cannot occur until (1)~all previous inputs
have occurred and (2)~either $i<N$ or else output number $i-N$ has
occurred.  Output number $i$ cannot occur until (1)~all previous
outputs have occurred and (2)~input number $i$ has occurred.  These
dependencies are illustrated pictorially in Figure~\ref{equivsketch1},
where circles represent the actions to be taken and arrows represent
dependency relationships between actions.

\begin{figure}[htbp]
\centerline{\epsfig{file=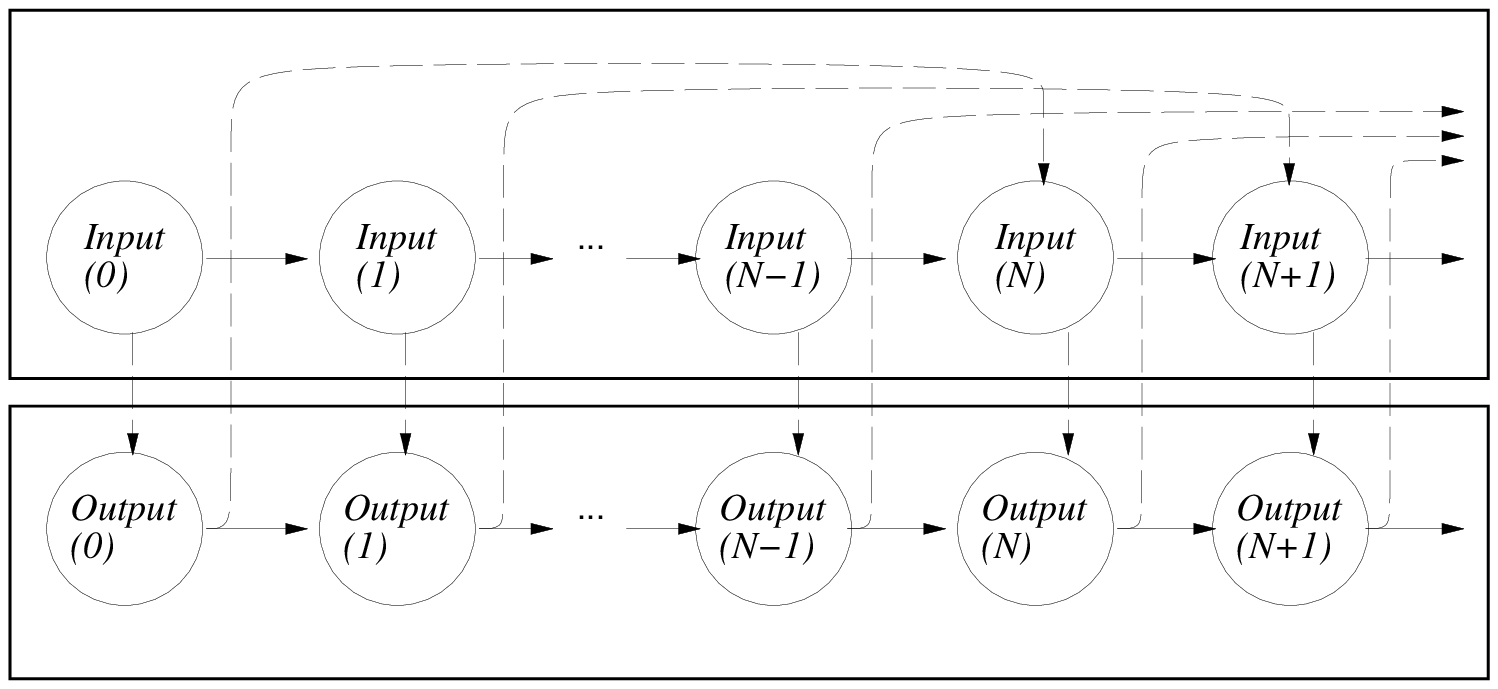,width=.8\textwidth}}
\caption{Moves of \RL.}\label{equivsketch2}
\end{figure}

\begin{figure}[htbp]
\centerline{\epsfig{file=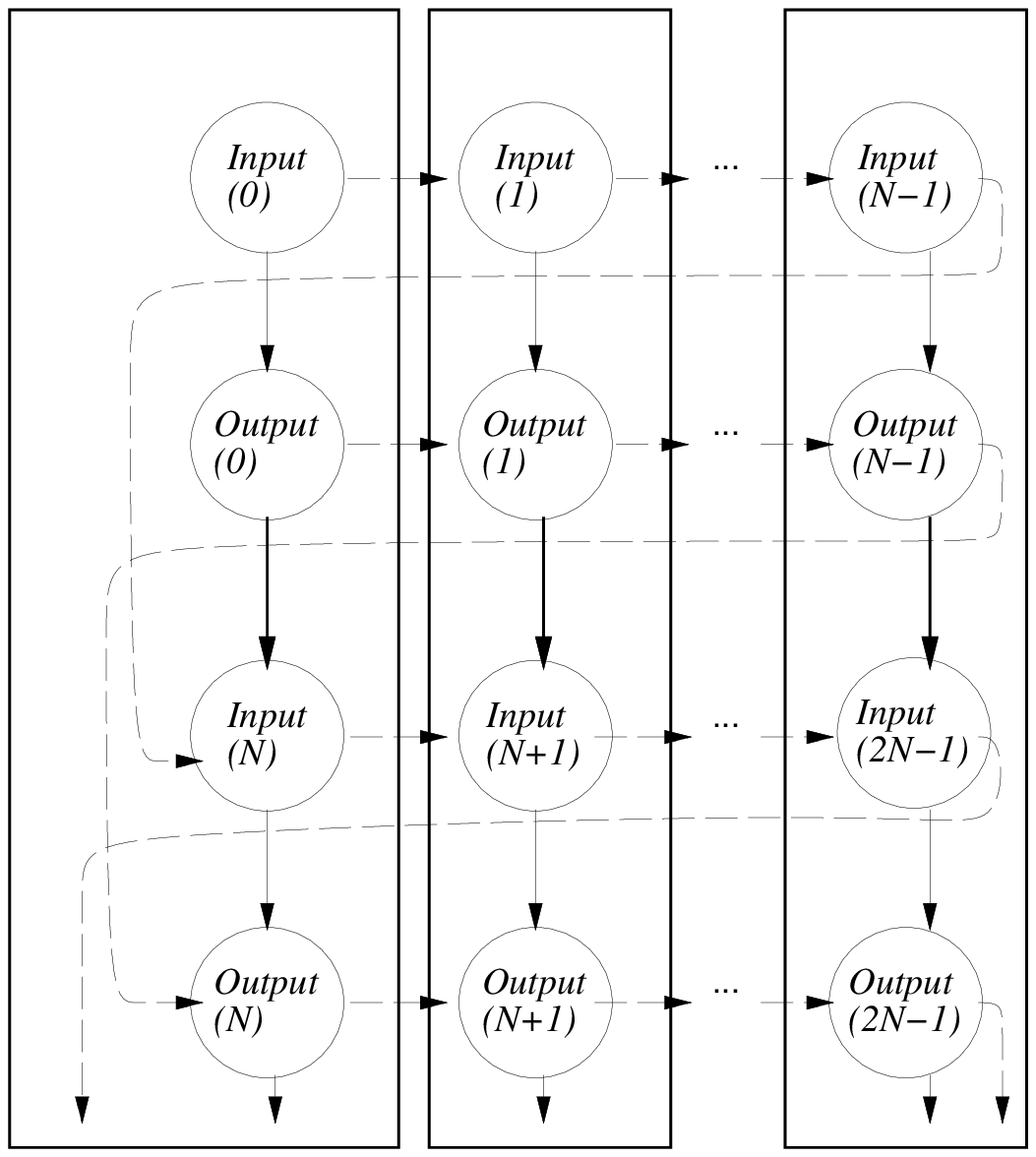,height=.72\textwidth}}
\caption{Moves of \CL.}\label{equivsketch3}
\end{figure}

Lamport writes the two programs in a semi-formal language reminiscent
of CSP \cite{csp} which we call Pseudo-CSP.  The first program, which
we denote by \RL, is shown in Figure~\ref{Rpcsp}.  It operates the
buffer using two processes; one handles input into the buffer
and the other handles output from the buffer.  It gives rise to a
row-wise decomposition of the graph of moves, as shown in
Figure~\ref{equivsketch2}.  The second program, which we denote by \CL, is
shown in Figure~\ref{Cpcsp}.  It uses $N$ processes, each 
managing input and output for one particular slot in the buffer.  It
gives rise to a column-wise decomposition of the graph of moves, as
shown in Figure~\ref{equivsketch3}.

\begin{figure}[htbp]
\begin{center}
\framebox{
$\begin{array}{l}
\mbox{\em in,\ out} : \mbox{\bf channel\ of\ } \mbox{\em Value}\\
\mbox{\em buf} : \mbox{\bf array\ } 0 \ldots N-1 \mbox{\bf\ of\ }
\mbox{\em Value}\\
p,g : \mbox{\bf internal\ } \mbox{\em Natural\ } \mbox{\bf initially\ } 0\\
\begin{array}{ll}
\mbox{\em Receiver} :: & * 
	\left[
	\begin{array}{rcl}
	p - g \neq N  & \rightarrow & 
			\mbox{\em in\ } ? \mbox{\em buf\ } [p \bmod N];\\
	 & & p := p + 1
	\end{array}
	\right] 
\\
\quad \| \\
\mbox{\em Sender} :: & * 
	\left[
	\begin{array}{rcl}
	p - g \neq 0  & \rightarrow & 
			\mbox{\em out\ } ! \mbox{\em buf\ } [g \bmod N];\\
	 & & g := g + 1
	\end{array}
	\right]	
\end{array} 
\end{array}$}
\end{center} 
\caption{A two-process ring buffer \RL, in Pseudo-CSP.}\label{Rpcsp}
\end{figure}

\begin{figure}[htbp]
\begin{center}
\framebox{
$\begin{array}{l}
\mbox{\em in,\ out} : \mbox{\bf channel\ of\ } \mbox{\em Value}\\
\mbox{\em buf} : \mbox{\bf array\ } 0 \ldots N-1 \mbox{\ of\ }
				\mbox{\em Value}\\
pp,gg : \mbox{\bf internal\ array\ } 0 \ldots N-1 \mbox{\bf\ of\ } \{0,1\}
	\mbox{\bf \ initially\ } 0\\
\mbox{\em Buffer}(i: 0 \ldots N-1) ::\\
\quad * \left[
\begin{array}{rcl}
\mbox{\em empty}:\ \mbox{\em IsNext}(pp,i) & \rightarrow & 
	\mbox{\em in\ } ? \mbox{\em buf\ }[i]; \\
 & & pp[i] := (pp[i]+1) \bmod 2; \\
\mbox{\em full}:\ \mbox{\em IsNext}(gg,i) & \rightarrow & 
	\mbox{\em out\ } ! \mbox{\em buf\ }[i]; \\
 & & gg[i] := (gg[i]+1) \bmod 2;
\end{array}
\right]  \\
\\
\begin{array}{rllll}
\mbox{\em IsNext}(r,i) & \stackrel{\triangle}= & \mbox{\bf if\ }\ i=0 &
	\mbox{\bf\ then\ } & r[0] = r[N-1]\\
 & &  & \mbox{\bf\ else\ } & r[i] \neq r[i-1]
\end{array}
\end{array}$
} 
\end{center}
\caption{An $N$ process ring buffer \CL, in Pseudo-CSP.}\label{Cpcsp}
\end{figure}

In Pseudo-CSP, the semicolon represents sequential composition, $\|$
represents parallel composition, and $*$ represents iteration.  The
general meanings of ? and ! are more complicated; they indicate
synchronization.  In the context of \RL\ and \CL, ``in ?'' is
essentially a command to place the current input into the given slot,
and ``out !'' is essentially a command to send out the datum in the
given slot as an output.  In Section~\ref{ringeas}, we
will give a more complete explanation of the two programs in terms of
evolving algebras.

After presenting the two algorithms in Pseudo-CSP, Lamport describes
them by means of formulas in TLA, the Temporal Logic of Actions
\cite{tla}, and proves the equivalence of the two formulas in TLA.  He
does not prove that the TLA formulas are equivalent to the
corresponding Pseudo-CSP programs.  The Pseudo-CSP presentations are
there only to guide the reader's intuition.  As we have mentioned,
Pseudo-CSP is only semi-formal; neither the syntax nor the semantics
of it is given precisely.

However, Lamport provides a hint as to why the two programs themselves
are equivalent.  There is a close correspondence of values between $p$
and $pp$, and between $g$ and $gg$.  Figure~\ref{pptable}, taken from
\cite{lamport}, illustrates the correspondence between $p$ and $pp$
for $N=4$.  The $n$th row describes the values of variables $p$ and
$pp$ after $n$ inputs.  The predicate IsNext(pp,i) is intended to
be true only for one array position $i$ at any state (the position
that is going to be active); the box
indicates that position.

\begin{figure}[htbp]
\begin{center}
\framebox{
$
\begin{array}{cccccc}
p & & pp[0] & pp[1] & pp[2] & pp[3] \\
\\
0 & & \fbox{0} & 0 & 0 & 0 \\
1 & & 1 & \fbox{0} & 0 & 0 \\
2 & & 1 & 1 & \fbox{0} & 0 \\
3 & & 1 & 1 & 1 & \fbox{0} \\
4 & & \fbox{1} & 1 & 1 & 1 \\
5 & & 0 & \fbox{1} & 1 & 1 \\
6 & & 0 & 0 & \fbox{1} & 1 \\
\vdots & & \vdots & \vdots & \vdots & \vdots\\
\end{array}
$
}
\end{center}
\caption{The correspondence between values of $pp$
and $p$, for $N=4$.}\label{pptable}
\end{figure}

\subsection{Discussion}

There are three issues where we disagree with Lamport.

\paragraph{Issue 1: The Notion of Equivalence.}

What does it mean that two programs are equivalent?  In our opinion,
the answer to the question depends on the desired abstraction
\cite{chall}.  There are many reasonable definitions of equivalence.
Here are some examples.

\begin{enumerate}
\item The two programs produce the same output
on the same input.

\item The two programs produce the same output on
the same input, and the two programs are of the same time
complexity (with respect to your favorite definition of
time complexity). \label{ordertime}

\item Given the same input, the two programs produce the same output
and take {\em precisely\/} the same amount of time. \label{sametime}

\item No observer of the execution of the two programs can detect
any difference.  \label{obs}

\end{enumerate}
The reader will be able to suggest numerous other reasonable
definitions for equivalence.  For example, one could substitute space
for time in conditions (\ref{ordertime}) and (\ref{sametime}) above.   The
nature of an ``observer'' in condition (\ref{obs}) admits
different plausible interpretations, depending upon what aspects
of the execution the observer is allowed to observe.  

Let us stress that we do not promote any particular notion of
equivalence or any particular class of such notions.  We only note
that there are different reasonable notions of equivalence and there
is no one notion of equivalence that is best for all purposes.  The
two ring-buffer programs are indeed ``strongly equivalent''; in
particular, they are equivalent in the sense of definition
(\ref{sametime}) above. However, they are not equivalent in the sense of
definition (\ref{obs}) for certain observers, or in the sense of some
space-complexity versions of definitions (\ref{ordertime}) and
(\ref{sametime}).  See Section~\ref{inequiv} in this connection.

\paragraph{Issue 2: Representing Programs as Formulas.}

Again, we quote Lamport \cite{lamport}:

\begin{quote} \begin{em}
We will not attempt to give a rigorous meaning to the program
text. Programming languages evolved as a method of describing
algorithms to compilers, not as a method for reasoning about them. We
do not know how to write a completely formal proof that two
programming language representations of the ring buffer are
equivalent.  In Section~2, we represent the program formally in TLA,
the Temporal Logic of Actions \cite{tla}.
\end{em} \end{quote}

We believe that it is not only possible but also beneficial to give a
rigorous meaning to one's programming language and to prove the
desired equivalence of programs directly.  The evolving algebra method
has been used to give rigorous meaning to various programming
languages \cite{biblio,www}.  In a similar way, one may try to give formal
semantics to Pseudo-CSP (which is used in fact for describing
algorithms to humans, not compilers).  Taking into account the
modesty of our goals in this paper, we do not do that and represent
\RL\ and \CL\ directly as evolving algebra programs \Rea\ and \Cea\
and then work with the two evolving algebras.

One may argue that our translation is not perfectly faithful.  Of
course, no translation from a semi-formal to a formal language can be
proved to be faithful.  We believe that our translation is reasonably
faithful; we certainly did not worry about the complexity of our
proofs as we did our translations.  Also, we do not think that
Lamport's TLA description of the Pseudo-CSP is perfectly faithful (see
the discussion in subsection~3.2) and thus we have two slightly different
ideals to which we can be faithful.  In fact, we do not think that
perfect faithfulness is crucially important here.  We give two
programming language representations \Rea\ and \Cea\ of the ring
buffer reflecting different decompositions of the buffer into
processes.  Confirming Lamport's thesis, we prove that the two
programs are equivalent in a very strong sense; our equivalence proof
is direct.  Then we point out that our programs are inequivalent according
to some natural definitions of equivalence.  Moreover, the same
inequivalence arguments apply to \RL\ and \CL\ as well.

\paragraph{Issue 3: The Formality of Proofs.}

Continuing, Lamport writes \cite{lamport}:

\begin{quote} \begin{em}
We now give a hierarchically structured proof that $\Pi_2$ and $\Pi_N$
[the TLA translations of \RL\ and \CL\ -- GH]
are equivalent \cite{proof}.  The proof is completely formal, meaning
that each step is a mathematical formula. English is used only to
explain the low-level reasoning. The entire proof could be carried
down to a level at which each step follows from the simple application
of formal rules, but such a detailed proof is more suitable for
machine checking than human reading. Our complete proof, with ``Q.E.D.''
steps and low-level reasoning omitted, appears in Appendix A.
\end{em} \end{quote}

We prefer to separate the process of explaining a proof to people from
the process of computer-aided verification of the same proof
\cite{beatcs}.  A human-oriented exposition is much easier for humans
to read and understand than expositions attempting to satisfy both
concerns at once.  Writing a good human-oriented proof is the art of
creating the correct images in the mind of the reader.  Such a proof
is amenable to the traditional social process of debugging
mathematical proofs.

Granted, mathematicians make mistakes and computer-aided verification
may be desirable, especially in safety-critical applications.  In this
connection we note that a human-oriented proof can be a starting point
for mechanical verification.  Let us stress also that a human-oriented
proof need not be less precise than a machine-oriented proof; it simply
addresses a different audience.

\paragraph{Revisiting Lamport's Thesis}
These disagreements do not mean that our position on ``the
insubstantiality of processes'' is the direct opposite of Lamport's.
We simply point out that ``the insubstantiality of processes'' may
itself be in the eye of the beholder.  The same two programs can be
equivalent with respect to some reasonable definitions of equivalence
and inequivalent with respect to others.

\section{Evolving Algebras}
\label{eaintro}

Evolving algebras were introduced in \cite{tutorial}; a more
detailed definition has appeared in
\cite{guide}.  Since its introduction, this methodology has been used for a
wide variety of applications: programming language semantics, hardware
specification, protocol verification, {\em etc.}.
It has been used to show equivalences of various kinds,
including equivalences across a variety of abstraction levels for
various real-world systems, {\em e.g.\/} \cite{wam}.  
See \cite{biblio,www} for numerous other examples.

We recall here only as much of evolving algebra definitions
\cite{guide} as needed in this
paper.  Evolving algebras (often abbreviated {\em ealgebras\/} or {\em
EA\/}) have many other capabilities not shown here: for example,
creating or destroying agents during the evolution.

Those already familiar with ealgebras may wish to skip this section.

\subsection{States}

States are essentially logicians' structures except that relations are
treated as special functions.  They are also called {\em static
algebras\/} and indeed they are algebras in the sense of the science
of universal algebra.

A {\em vocabulary\/} is a finite collection of function names, each of
fixed arity.  Every vocabulary contains the following {\em logic
symbols\/}: nullary function names {\em true, false, undef}, the
equality sign, (the names of) the usual Boolean operations and (for
convenience) a unary function name Bool.  Some function symbols are
tagged as relation symbols (or predicates); for example, Bool and the
equality sign are predicates.

A {\em state\/} $S$ {\em of vocabulary\/} $\Upsilon$ is a non-empty set $X$
(the {\em basic set\/} or {\em superuniverse\/} of $S$), together with
interpretations of all function symbols in $\Upsilon$ over $X$ (the {\em
basic functions\/} of $S$).  A function symbol $f$ of arity $r$ is
interpreted as an $r$-ary operation over $X$ (if $r=0$, it is
interpreted as an element of $X$).  The interpretations of predicates
(the {\em basic relations\/}) and the logic symbols satisfy the
following obvious requirements.  The elements (more exactly, the
interpretations of) {\em true\/} and {\em false\/} are distinct.
These two elements are the only possible values of any basic relation
and the only arguments where Bool produces {\em true}.  They are
operated upon in the usual way by the Boolean operations.  The
interpretation of {\em undef\/} is distinct from those of {\em true\/}
and {\em false}.  The equality sign is interpreted as the equality
relation.  We denote the value of a term $t$ in state $S$ by $t_S$.

Domains.  Let $f$ be a basic function of arity $r$ and $\bar x$ range
over $r$-tuples of elements of $S$.  If $f$ is a basic relation then
the {\em domain of $f$\/} at $S$ is $\{\bar x: f(\bar x) = \mbox{true}\}$.
Otherwise the {\em domain of $f$\/} at $S$ is $\{\bar x: f(\bar x) \neq
\mbox{undef}\}$.

Universes. A basic relation $f$ may be viewed as the set of tuples
where it evaluates to {\em true}.  If $f$ is unary it can be viewed as
a {\em universe}.  For example, Bool is a universe consisting of
two elements (named) {\em true\/} and {\em false}.  Universes allow
us to view states as many-sorted structures.

Types.  Let $f$ be a basic function of arity $r$ and $U_0,\ldots,U_r$
be universes.  We say that $f$ is {\em of type\/} $U_1\times
\cdots\times U_r\rightarrow U_0$ in the given state if the domain of
$f$ is $U_1\times \cdots\times U_r$ and $f(\bar x)\in U_0$ for every
$\bar x$ in the domain of $f$.  In particular, a nullary $f$ is of type
$U_0$ if (the value of) $f$ belongs to $U_0$.

Example.  Consider a directed ring of nodes with two tokens; each node
may be colored or uncolored.  We formalize this as a state as follows.
The superuniverse contains a non-empty universe Nodes comprising the
nodes of the ring.  Also present is the obligatory two-element
universe Bool, disjoint from Nodes.  Finally, there is an element
(interpreting) {\em undef\/} outside of Bool and outside of Nodes.
There is nothing else in the superuniverse.  (Usually we skip the
descriptions of Bool and {\em undef\/}).  A unary function Next
indicates the successor to a given node in the ring.  Nullary
functions Token1 and Token2 give the positions of the two tokens.  A
unary predicate Colored indicates whether the given node is colored.

\subsection{Updates}

There is a way to view states which is unusual to logicians.  
View a state as a sort of memory.  Define a {\em
location\/} of a state $S$ to be a pair $\ell=(f,\bar x)$, where $f$ is
a function name in the vocabulary of $S$ and $\bar x$ is a tuple of
elements of (the superuniverse of) $S$ whose length equals the arity
of $f$.  (If $f$ is nullary, $\ell$ is simply $f$.)  In the two-token
ring example, let $a$ be any node (that is, any element of the
universe Nodes).  Then the pair (Next,$a$) is a location.

An {\em update\/} of a state $S$ is a pair $\alpha=(\ell,y)$, where
$\ell$ is a location of $S$ and $y$ is an element of $S$.  To {\em
fire\/} $\alpha$ at $S$, put $y$ into the location $\ell$; that is, if
$\ell=(f,\bar x)$, redefine $S$ to interpret $f(\bar x)$ as $y$; nothing
else (including the superuniverse) is changed.  We say that an update
$(\ell,y)$ of state $S$ is {\em trivial\/} if $y$ is the content of
$\ell$ in $S$.  In the two-token ring example, let $a$ be any node.
Then the pair (Token1, $a$) is an update.  To fire this update, move
the first token to the position $a$.

Remark to a curious reader.  If $\ell$ = (Next,$a$), then ($\ell,a$)
is also an update.  To fire this update, redefine the successor of
$a$; the new successor is $a$ itself.  This update destroys the ring
(unless the ring had only one node).  To guard from such undesirable
changes, the function Next can be declared static (see \cite{guide})
which will make any update of Next illegal.

An {\em update set\/} over a state $S$ is a set of updates of $S$.
An update set is {\em consistent\/} at $S$ if no two updates
in the set have the same location but different values.  To
fire a consistent set at $S$, fire all its members simultaneously;
to fire an inconsistent set at $S$, do nothing.
In the two-token ring example, let $a,b$ be two nodes. Then the update
set $\{(Token1,a),(Token1,b)\}$ is consistent if and only if $a=b$.

\subsection{Basic Transition Rules}

We introduce rules for changing states.  The semantics for each rule
should be obvious.  At a given state $S$ whose vocabulary includes
that of a rule $R$, $R$ gives rise to an update set $\US(R,S)$; to
execute $R$ at $S$, one fires $\US(R,S)$.  We say that $R$ is {\em enabled\/}
at $S$ if $\US(R,S)$ is consistent and contains a non-trivial update.
We suppose below that a state of discourse $S$ has a sufficiently rich
vocabulary.

An {\em update instruction\/} $R$ has the form

\begin{minipage}{\textwidth}
\begin{tabbing} \qquad\=\qquad\=\kill
\> $f(t_1, \ldots, t_r) := t_0$ 
\end{tabbing}
\end{minipage}

\noindent where $f$ is a function name of arity $r$ and each $t_i$ is
a term.  (If $r=0$ we write ``$f:=t_0$'' rather than ``$f():=t_0$''.)
The update set $\US(R,S)$ contains a single element
$(\ell,y)$, where $y$ is the value $(t_0)_S$ of $t_0$ at $S$ and $\ell
= (f, (x_1, \ldots, x_r))$ with $x_i = (t_i)_S$.  In other words, to
execute $R$ at $S$, set $f((t_1)_S,\ldots,(t_r)_S)$ to $(t_0)_S$ and
leave the rest of the state unchanged.  In the two-token ring example,
``Token1 := Next(Token2)'' is an update instruction.  To execute it, move
token 1 to the successor of (the current position of) token 2. 

A {\em block rule\/} $R$ is a sequence $R_1,\ldots,R_n$ of transition
rules.  To execute $R$ at $S$, execute all the constituent rules at
$S$ simultaneously.  More formally, $\US(R,S) = \bigcup_{i=1}^{n}
\US(R_i,S)$. (One is supposed to write ``{\bf block}'' and ``{\bf
endblock}'' to denote the scope of a block rule; we often omit them
for brevity.)  In the two-token ring example, consider the following
block rule:

\begin{minipage}{\textwidth}
\begin{tabbing}\qquad\=\qquad\=\kill
\> Token1 := Token2\\
\> Token2 := Token1
\end{tabbing}
\end{minipage}

\noindent
To execute this rule, exchange the tokens.  The new position of
Token1 is the old position of Token2, and the new position of Token2
is the old position of Token1.  

A {\em conditional rule\/} $R$ has the form

\begin{minipage}{\textwidth}
\begin{tabbing} \qquad\=\qquad\=\kill
\> \If $g$ \Then $R_0$ \Endif 
\end{tabbing}
\end{minipage}

\noindent where $g$ (the {\em guard}) is a term and $R_0$ is a rule.  
If $g$ holds (that is, has the same value as {\em true\/}) in $S$ then
$\US(R,S) = \US(R_0,S)$; otherwise $\US(R,S) = \emptyset$.  (A more
general form is ``\If $g$ \Then $R_0$ \Else $R_1$ \Endif'', but we do
not use it in this paper.)  In the two-token ring example, consider
the following conditional rule:

\begin{minipage}{\textwidth}
\begin{tabbing} \qquad\=\qquad\=\kill
\> \If Token1 = Token2 \Then\\
\> \> Colored(Token1) := true\\
\> \Endif
\end{tabbing}
\end{minipage}

\noindent
Its meaning is the following: if the two tokens are at the same node,
then color that node.

\subsection{Rules with Variables}

Basic rules are sufficient for many purposes, e.g. to give operational
semantics for the C programming language \cite{c}, but in this paper
we need two additional rule constructors.  The new rules use
variables.  Formal treatment of variables requires some care but the
semantics of the new rules is quite obvious, especially
because we do not need to nest constructors with variables here.  Thus
we skip the formalities and refer the reader to \cite{guide}.  As
above $S$ is a state of sufficiently rich vocabulary.

A {\em parallel synchronous rule\/} (or {\em declaration rule},
as in \cite{guide}) $R$ has the form:

\begin{minipage}{\textwidth}
\begin{tabbing} \qquad\=\qquad\=\qquad\=\kill
\> \Var $x$ \RangesOver $U$\\
\> \> $R(x)$\\
\> \Endvar
\end{tabbing}
\end{minipage}

\noindent 
where $x$ is a variable name, $U$ is a universe name, and $R(x)$ can
be viewed as a rule template with free variable $x$.  To execute $R$
at $S$, execute simultaneously all rules $R(u)$ where $u$ ranges over
$U$.  In the two-token ring example, (the execution of)
the following rule colors all nodes except for the nodes occupied by
the tokens.

\begin{minipage}{\textwidth}
\begin{tabbing}\qquad\=\qquad\=\qquad\=\kill
\> \Var $x$ \RangesOver Nodes\\
\> \> \If $x \neq$ Token1 \And $x \neq$ Token2 \Then\\
\> \> \> Colored(x) := true\\
\> \> \Endif\\
\> \Endvar
\end{tabbing}
\end{minipage}

A {\em choice rule\/} $R$ has the form

\begin{minipage}{\textwidth}
\begin{tabbing}\qquad\=\qquad\=\qquad\=\kill
\> \Choose $x$ \In $U$\\
\> \> $R(x)$\\
\> \Endchoose
\end{tabbing}
\end{minipage}

\noindent 
where $x$, $U$ and $R(x)$ are as above.  It is nondeterministic.  To
execute the choice rule, choose arbitrarily one element $u$ in $U$ and
execute the rule $R(u)$.  In the two-token ring example, each
execution of the following rule either colors an unoccupied node or
does nothing.

\begin{minipage}{\textwidth}
\begin{tabbing}\qquad\=\qquad\=\qquad\=\kill
\> \Choose $x$ \In Nodes\\
\> \> \If $x \neq$ Token1 \And $x \neq$ Token2 \Then\\
\> \> \> Colored(x) := true\\
\> \> \Endif\\
\> \Endchoose
\end{tabbing}
\end{minipage}
  
\subsection{Distributed Evolving Algebra Programs}

Let $\Upsilon$ be a vocabulary that contains the universe {\em Agents}, the
unary function {\em Mod\/} and the nullary function {\em Me}.
A {\em distributed EA program\/} $\Pi$ of vocabulary
$\Upsilon$ consists of a finite set of {\em modules}, each of which is
a transition rule with function names from $\Upsilon$.  Each module is
assigned a different name; these names are nullary function names from
$\Upsilon$ different from {\em Me}.  Intuitively, a module is the
program to be executed by one or more agents.

A (global) {\em state\/} of $\Pi$ is a structure $S$ of vocabulary
$\Upsilon$--\{Me\} where different module names are interpreted as
different elements of $S$ and the function {\em Mod\/} assigns (the
interpretations of) module names to elements of {\em Agents}; {\em
Mod\/} is undefined (that is, produces {\em undef\/}) otherwise.  If
{\em Mod\/} maps an element $\alpha$ to a module name $M$, we say that
$\alpha$ is an {\em agent\/} with program $M$.

For each agent $\alpha$, View$_\alpha(S)$ is the reduct of $S$ to the
collection of functions mentioned in the module Mod($\alpha$),
expanded by interpreting {\em Me\/} as $\alpha$. Think about
View$_\alpha(S)$ as the local state of agent $\alpha$ corresponding to
the global state $S$.  We say that an agent $\alpha$ is {\em enabled\/} at
$S$ if Mod($\alpha$) is enabled at View$_\alpha(S)$; that is, if the
update set generated by Mod($\alpha$) at View$_\alpha(S)$ is
consistent and contains a non-trivial update.  This update set is also
an update set over $S$.  To {\em fire\/} $\alpha$ at $S$, execute that
update set.

\subsection{Runs}
\label{runs}

In this paper, agents are not created or destroyed.  Taking this into
account, we give a slightly simplified definition of runs.

A {\em run\/} $\rho$ of a distributed ealgebra program $\Pi$ of vocabulary
$\Upsilon$ from the initial state $S_0$ is a triple $(M, A,
\sigma)$ satisfying the following conditions.

\begin{description}
\item[1.] $M$, the set of {\em moves\/} of $\rho$, is a partially ordered
set where every $\{\nu: \nu \leq\mu\}$ is finite.  

Intuitively, $\nu <\mu$ means that move $\nu$ completes before move
$\mu$ begins. If $M$ is totally ordered, we say that $\rho$ is a {\em
sequential\/} run.

\item[2.] $A$ assigns agents (of $S_0$) to moves in such a way that every
non-empty set $\{\mu: A(\mu) = \alpha\}$ is linearly ordered.  

Intuitively, $A(\mu)$ is the agent performing move $\mu$; every agent acts
sequentially.

\item[3.] $\sigma$ maps finite initial segments of $M$ (including $\emptyset$)
to states of $\Pi$.

Intuitively, $\sigma(X)$ is the result of performing
all moves of $X$; $\sigma(\emptyset)$ is the initial state $S_0$.  States
$\sigma(X)$ are the {\em states of $\rho$}.

\item[4.] {\em Coherence}.  If $\mu$ is a maximal element of a finite
initial segment $Y$ of $M$, and $X = Y - \{\mu\}$, then $A(\mu)$ is
enabled at $\sigma(X)$ and $\sigma(Y)$ is obtained by firing $A(\mu)$ at
$\sigma(X)$.

\end{description}

It may be convenient to associate particular states with single moves.
We define $\Lambda(\mu) = \sigma(\{\nu: \nu < \mu\})$.

The definition of runs above allows no interaction between the agents
on the one side and the external world on the other.  In such a case,
a distributed evolving algebra is given by a program and the collection
of initial states.  In a more general case, the environment can
influence the evolution.  Here is a simple way to handle interaction
with the environment which suffices for this paper.

Declare some basic functions (more precisely, some function names)
{\em external}.  Intuitively, only the outside world can change them.
If $S$ is a state of $\Pi$ let $S^-$ be the reduct of $S$ to (the
vocabulary of) non-external functions.  Replace the coherence
condition with the following:

\begin{description}

\item[4$'$.] {\em Coherence}.  If $\mu$ is a maximal element of a
finite initial segment $Y$ of $M$, and $X = Y - \{\mu\}$, then
$A(\mu)$ is enabled in $\sigma(X)$ and $\sigma(Y)^-$ is obtained by firing
$A(\mu)$ at $\sigma(X)$ and forgetting the external functions.
\end{description}

In applications, external functions usually satisfy certain
constraints.  For example, a nullary external function Input may
produce only integers.  To reflect such constraints, we define {\em
regular runs\/} in applications.  A distributed evolving algebra is
given by a program, the collection of initial states and the collection of
regular runs.  (Of course, regular runs define the initial states, but
it may be convenient to specify the initial states separately.)

\section{The Ring Buffer Evolving Algebras}
\label{ringeas}

The evolving algebras \Rea and \Cea, our ``official'' representations
of \RL\ and \CL, are given in subsections~\ref{officialRea} and
\ref{officialCea}; see Figures~\ref{Rea} and \ref{Cea}.  The reader
may proceed there directly and ignore the preceding subsections where
we do the following.  We first present in subsection~\ref{R1sect} an
elaborate ealgebra R1 that formalizes \RL\ together with its
environment; R1 expresses our understanding of how \RL\ works, how it
communicates with the environment and what the environment is supposed
to do.  Notice that the environment and the synchronization magic of
CSP are explicit in R1.  In subsection~\ref{R2sect}, we then transform
R1 into another ealgebra R2 that performs synchronization implicitly.
We transform R2 into \Rea\ by parallelizing the rules slightly and
making the environment implicit; the result is shown in
subsection~\ref{officialRea}.  (In a sense, R1, R2, and \Rea\ are all
equivalent to another another, but we will not formalize this.) 
We performed a similar analysis and transformation to create \Cea\
from \CL; we omit the intermediate stages and present \Cea\ directly
in subsection~\ref{officialCea}.  

\subsection{R1: The First of the Row Evolving Algebras}
\label{R1sect}

The program for R1, given in Figure~\ref{R1}, contains six modules.
The names of the modules reflect the intended meanings.  In particular,
modules BuffFrontEnd and BuffBackEnd correspond to the two processes
Receiver and Sender of \RL.

\begin{figure}[htbp]
\begin{tabbing}
\quad\=\qquad\=\qquad\=\kill
\rule{\textwidth}{1pt}\\

Module InputEnvironment\\
\>\If Mode(Me) = Work \Then\\
\> \>\Choose $v$ \In Data\\
\> \> \> InputDatum := $v$\\
\> \>\Endchoose\\
\> \> Mode(Me) := Ready\\
\>\Endif\\ 
\rule{\textwidth}{1pt}\\

Module OutputEnvironment\\
\>\If Mode(Me) = Work \Then Mode(Me) := Ready \Endif\\
\rule{\textwidth}{1pt}\\

Module InputChannel\\
\>\If Mode(Sender(Me)) = Ready \And Mode(Receiver(Me)) = Ready \Then\\
\> \> Buffer($p \bmod N$) := InputDatum\\
\> \> Mode(Sender(Me))    := Work\\
\> \> Mode(Receiver(Me))  := Work\\
\>\Endif\\
\rule{\textwidth}{1pt}\\

Module OutputChannel\\
\>\If Mode(Sender(Me)) = Ready \And Mode(Receiver(Me)) = Ready \Then\\
\> \> OutputDatum := Buffer($g \bmod N$)\\
\> \> Mode(Sender(Me))    := Work\\
\> \> Mode(Receiver(Me))  := Work\\
\>\Endif\\
\rule{\textwidth}{1pt}\\

Module BuffFrontEnd\\	
Rule FrontWait\\
\>\If Mode(Me) = Wait \And $p - g \neq N$ \Then Mode(Me) := Ready \Endif\\
Rule FrontWork\\
\>\If Mode(Me) = Work \Then $p$ := $p + 1$, Mode(Me) := Wait \Endif\\
\rule{\textwidth}{1pt}\\

Module BuffBackEnd\\
Rule BackWait\\
\>\If Mode(Me) = Wait \And $p - g \neq 0$ \Then Mode(Me) := Ready \Endif\\
Rule BackWork\\
\>\If Mode(Me) = Work \Then $g$ := $g + 1$, Mode(Me) := Wait \Endif\\
\rule{\textwidth}{1pt}
\end{tabbing}
\caption{The program for R1.} \label{R1}
\end{figure}

Comment for ealgebraists.  In terms of \cite{guide}, the
InputChannel agent is a two-member team comprising the
InputEnvironment and the BuffFrontEnd agents; functions Sender and
Receiver are similar to functions Mem\-ber$_1$ and Mem\-ber$_2$.
Similarly the OutputChannel agent is a team.  This case is very simple
and one can get rid of unary functions Sender and Receiver by
introducing names for the sending and receiving agents.

Comment for CSP experts.  Synchronization is implicit in CSP.  It is a
built-in magic of CSP.  We have doers of synchronization.  (In this
connection, the reader may want to see the EA treatment of Occam in
\cite{occam}.)  Nevertheless, synchronization remains abstract.  In a
sense the abstraction level is even higher: similar agents can
synchronize more than two processes.

Comment.  The nondeterministic formalizations of the input and output
environments are abstract and may be refined in many ways.

\paragraph{Initial states.\ }
In addition to the function names mentioned in the program (and the logic
names), the vocabulary of R1 contains universe names Data, Integers,
\ZN, \Ztwo, Modes and a subuniverse Senders-and-Receivers of Agents.
Initial states of R1 satisfy the following requirements.
\begin{enumerate}

\item  The universe Integers and the arithmetical function names mentioned
in the program have their usual meanings.  The universe \ZN\ consists
of integers modulo $N$ identified with the integers $0,\ldots,N-1$.
The universe \Ztwo\ is similar.  $p = g = 0$.  Buffer is of type \ZN\
$\rightarrow$ Data; InputDatum and OutputDatum take values in Data.

\item The universe Agents contains six elements to which Mod assigns
different module names.  We could have special nullary functions to
name the six agents but we don't; we will call them with respect to
their programs: the input environment, the output environment, the
input channel, the output channel, buffer's front end and buffer's
back end respectively. Sender(the input channel) = the input
environment, Receiver(the input channel) = buffer's front end,
Sender(the output channel) = buffer's back end, and Receiver(the
output channel) = the output environment.  The universe
Senders-and-Receivers consists of the two buffer agents and the two
environment agents.  Nullary functions Ready, Wait and Work are
distinct elements of the universe Modes.  The function Mode is defined
only over Senders-and-Receivers.  For the sake of simplicity of
exposition, we assign particular initial values to Mode: it assigns
Wait to either buffer agent, Work to the input environment agent, and
Ready to the output environment agent.

\end{enumerate}

\paragraph{Analysis}
In the rest of this subsection, we prove that R1 has the intended
properties.

\begin{lem}[Typing Lemma for R1] In every state of any run of R1, the dynamic
functions have the following (intended) types.

\begin{enumerate}

\item Mode: Senders-and-Receivers $\rightarrow$ Modes.

\item InputDatum, OutputDatum: Data.

\item $p, g$: Integers.

\item Buffer: \ZN\ $\rightarrow$ Data. 

\end{enumerate}
\end{lem}

\begin{pf} By induction over states. \qed \end{pf}

\begin{lem}[The p and g Lemma for R1]
Let $\rho$ be an arbitrary run of R1.  In every state of $\rho$, $0 \leq p
- g \leq N$.  Furthermore, if $p - g = 0$ then Mode(buffer's back end)
= Wait, and if $p - g = N$ then Mode(buffer's front end) = Wait.
\end{lem}

\begin{pf} 
An obvious induction.  See Lemma~\ref{lem1} in this regard.
\qed \end{pf}

\begin{lem}[Ordering Lemma for R1]
\label{orderingR1}
In any run of R1, we have the following.

\begin{enumerate}
\item 
If $\mu$ is a move of the input channel and $\nu$ is a move of
buffer's front end then either $\mu < \nu$ or $\nu < \mu$.

\item 
If $\mu$ is a move of the output channel and $\nu$ is a move of
buffer's back end then either $\mu < \nu$ or $\nu < \mu$.

\item 
For any buffer slot $k$, if $\mu$ is a move of the input channel
involving slot $k$ and $\nu$ is a move of the output channel involving
slot $k$ then either $\mu < \nu$ or $\nu < \mu$.
\end{enumerate}
\end{lem}

\begin{pf}
Let $\rho = (M,A,\sigma)$ be a run of R1.

\begin{enumerate}
\item \label{orderingpt1} 
Suppose by contradiction that $\mu$
and $\nu$ are incomparable and let $X = \{\pi : \pi < \mu
\vee \pi < \nu\}$ so that, by the coherence requirements on the run,
both agents are enabled at $\sigma(X)$, which is impossible because
their guards are contradictory.

Since the input channel is
enabled, the mode of buffer's front end is Ready at $X$.  But then
buffer's front end is disabled at $X$, which gives the desired
contradiction.

\item 
Similar to part (\ref{orderingpt1}).

\item 
Suppose by contradiction that $\mu$ and $\nu$ are incomparable and let
$X = \{\pi : \pi < \mu \vee \pi < \nu\}$ so that both agents are
enabled at $\sigma(X)$.  Since $\mu$ involves $k$, $p = k$ mod $N$ in
$\sigma(X)$.  Similarly, $g = k$ mod $N$ in $\sigma(X)$.  Hence $p - g
= 0$ mod $N$ in $\sigma(X)$.  By the p and g lemma, either $p - g = 0$
or $p - g = N$ in $\sigma(X)$.  In the first case, the mode of
buffer's back end is Wait and therefore the output channel is
disabled.  In the second case, the mode of buffer's front end is Wait
and therefore the input channel is disabled.  In either case, we have
a contradiction. \qed
\end{enumerate} \end{pf}

Recall that the state of move $\mu$ is $\Lambda(\mu) = \sigma(\{\nu :
\nu<\mu\})$.  By the coherence requirement, the agent $A(\mu)$ is enabled in 
$\Lambda(\mu)$.

Consider a run of R1.  Let $\mu_i$ (respectively, $\nu_i$) be the
$i$th move of the input channel (respectively, the output channel).
The value $a_i$ of InputDatum in $\Lambda(\mu_i)$ (that, is the datum
to be transmitted during $\mu_i$) is the {\em $i$th input datum}, and
the sequence $a_0,a_1,\ldots$ is the {\em input data sequence}.  (It
is convenient to start counting from $0$ rather than $1$.)  Similarly,
the value $b_j$ of OutputDatum in $\Lambda(\nu_j)$ is the {\em $j$th
output datum of $R$\/} and the sequence $b_0,b_1,\ldots$ is the {\em
output data sequence}.

Lamport writes:

\begin{quote} \begin{em}
To make the example more interesting, we assume no liveness properties
for sending values on the {\em in\/} channel, but we require that every
value received in the buffer be eventually sent on the {\em out\/}
channel.
\end{em} \end{quote}

With this in mind, we call a run {\em regular\/} if the output
sequence is exactly as long as the input sequence.

\begin{thm}
For a regular run, the output sequence is identical with the input sequence. 
\end{thm}

\begin{pf}
Let $\mu_0, \mu_1, \ldots$ be the moves of the input channel and
$\nu_0, \nu_1,\ldots$ be the moves of the output channel.  A simple
induction shows that $\mu_i$ stores the $i$th input datum $a_i$ at
slot $i \bmod N$ and $p = i$ at $\Lambda(\mu_i)$.  Similarly,
$\nu_j$ sends out the $j$th output datum $b_j$ from slot $j \bmod N$
and $g = j$ at $\Lambda(\nu_j)$.  If $\mu_i
< \nu_i < \mu_{i+N}$, then $a_i = b_i$.  We show that, for all $i$,
$\mu_i < \nu_i < \mu_{i+N}$.

By the $p$ and $g$ lemma, $p - g >0$ in $\Lambda(\nu_j)$ for any $j$,
and $p - g < N$ in $\Lambda(\mu_j)$ for any $j$.

\begin{enumerate}
\item
Suppose $\nu_i < \mu_i$.  Taking into account the monotonicity of
$p$, we have the following at $\Lambda(\nu_i)$: $p \leq i$, $g = i$
and therefore $p - g \leq 0$ which is impossible.

\item
Suppose $\mu_{i+N} < \nu_i$.  Taking into account the monotonicity
of $g$, we have the following at $\Lambda(\mu_{i+N})$: $p = i + N$, $g
\leq i $, and therefore $p - g \geq N$ which is impossible.
\end{enumerate}

By the ordering lemma, $\nu_i$ is order-comparable with both $\mu_i$
and $\mu_{i+N}$.  It follows that $\mu_i < \nu_i < \mu_{i+N}$.
\qed \end{pf}
  
\subsection{R2: The Second of the Row Evolving Algebras}
\label{R2sect}

One obvious difference between \RL\ and R1 is the following: R1
explicitly manages the communication channels between the buffer and
the environment, while \RL\ does not.  By playing with the modes of
senders and receivers, the channel modules of R1 provide explicit
synchronization between the environment and the buffers.  This
synchronization is implicit in the ``?'' and ``!'' operators of CSP.
To remedy this, we transform R1 into an ealgebra R2 in which
communication occurs implicitly.  R2 must somehow ensure
synchronization.  There are several options.

\begin{enumerate}
\item 
Allow BuffFrontEnd (respectively, BuffBackEnd) to modify the
mode of the input environment (respectively, the output environment)
to ensure synchronization.

This approach is feasible but undesirable.  It is unfair; the buffer
acts as a receiver on the input channel and a sender on the output
channel but exerts complete control over the actions of both
channels.  Imagine that the output environment represents another
buffer, which operates as our buffer does; in such a case both agents
would try to exert complete control over the common channel.

\item 
Assume that BuffFrontEnd (respectively, BuffBackEnd) does not
execute until the input environment (respectively, the output
environment) is ready.

This semantical approach reflects the synchronization magic of CSP.
It is quite feasible.  Moreover, it is common in the EA literature to
make assumptions about the environment when necessary.  It is not
necessary in this case because there are very easy programming
solutions (see the next two items) to the problem.

\item 
Use an additional bit for either channel which tells us whether the
channel is ready for communication or not.

In fact, a state of a channel comprises a datum and an additional bit
in the TLA part of Lamport's paper.  One can avoid dealing with states
of the channel by requiring that each sender and receiver across a
channel maintains its own bit (a well-known trick) which brings us to
the following option.

\item 
Use a bookkeeping bit for every sender and every receiver.
\end{enumerate}

It does not really matter, technically speaking, which of the four
routes is chosen.  To an extent, the choice is a matter of taste.  We
choose the fourth approach.  The resulting ealgebra R2 is shown in
Figure~\ref{R2}.

\begin{figure}[htbp]
\begin{tabbing}
\quad\=\qquad\=\qquad\=\kill
\rule{\textwidth}{1pt}\\

Module InputEnvironment\\
\>\If InSendBit = InReceiveBit Then\\
\> \>\Choose $v$ \In Data\\
\> \> \> InputDatum := $v$\\
\> \>\Endchoose\\
\> \> InSendBit := 1 -- InSendBit\\
\>\Endif\\ 
\rule{\textwidth}{1pt}\\

Module OutputEnvironment\\
\>\If OutSendBit $\neq$ OutReceiveBit \Then \\
\> \> OutReceiveBit := 1 -- OutReceiveBit\\
\>\Endif\\
\rule{\textwidth}{1pt}\\

Module BuffFrontEnd\\
Rule FrontWait\\
\>\If Mode(Me) = Wait \And $p - g \neq N$ \Then Mode(Me) := Ready \Endif\\
\\
Rule FrontCommunicate\\
\>\If Mode(Me) = Ready \And InSendBit $\neq$ InReceiveBit \Then\\
\> \> Buffer($p \bmod N$) := InputDatum\\
\> \> Mode(Me)            := Work\\
\> \> InReceiveBit := 1 -- InReceiveBit\\
\>\Endif\\
\\
Rule FrontWork\\
\>\If Mode(Me) = Work \Then $p$ := $p + 1$, Mode(Me) := Wait \Endif\\
\rule{\textwidth}{1pt}\\

Module BuffBackEnd\\
Rule BackWait\\
\>\If Mode(Me) = Wait \And $p - g \neq 0$ \Then Mode(Me) := Ready \Endif\\
\\
Rule BackCommunicate\\
\>\If Mode(Me) = Ready \And OutSendBit = OutReceiveBit \Then\\
\> \> OutputDatum := Buffer($g \bmod N$)\\
\> \> Mode(Me)    := Work\\
\> \> OutSendBit := 1 -- OutSendBit\\
\>\Endif\\
\\
Rule BackWork\\
\>\If Mode(Me) = Work \Then $g$ := $g + 1$, Mode(Me) := Wait \Endif\\
\rule{\textwidth}{1pt}
\end{tabbing}
\caption{The program for R2.} \label{R2}
\end{figure}

Notice that the sender can place data into a channel only when
the synchronization bits match, and the receiver can read the
data in a channel only when the synchronization bits do not match.

The initial states of R2 satisfy the first condition on the initial
states of R1.  The universe Agents contains four elements to which Mod
assigns different module names; we will call them with respect to
their programs: the input environment, the output environment,
buffer's front end, and buffer's back end, respectively.  The universe
BufferAgents contains the buffer's front end and buffer's back end
agents.  Nullary functions InSendBit, InReceiveBit, OutSendBit,
OutReceiveBit are all equal to $0$. Nullary functions Ready, Wait and
Work are distinct elements of the universe Modes.  The function Mode
is defined only over BufferAgents; it assigns Wait to each buffer
agent.  InputDatum and OutputDatum take values in Data.  Define the
input and output sequences and regular runs as in R1.

Let $\Upsilon_1$ be the vocabulary of R1 and $\Upsilon_2$ be the
vocabulary of R2.

\begin{lem}
Every run $R = (M,A,\sigma)$ of R1 induces a run $\rho = (M,B,\tau)$ of R2
where:

\begin{enumerate}
\item If $\mu \in M$ and $A(\mu)$ is not a channel agent, then $B(\mu)
= A(\mu)$. If $A(\mu)$ = the input channel, then $B(\mu)$ = buffer's
front end.  If $A(\mu)$ = the output channel, then $B(\mu)$ = buffer's
back end.

\item Let $X$ be a finite initial segment of $M$. 
$\tau(X)$ is the unique state satisfying the following conditions:
\begin{enumerate}

\item $\tau(X) | (\Upsilon_1 \cap \Upsilon_2) = 
	\sigma(X) | (\Upsilon_1 \cap \Upsilon_2)$

\item InReceiveBit = $p \bmod 2$ if the mode of buffer's front
end is Wait or Ready, and $1 - p \bmod 2$ otherwise.

\item OutSendBit = $g \bmod 2$ if the mode of buffer's back
end is Wait or Ready, and $1 - g \bmod 2$ otherwise.

\item InSendBit = InReceiveBit if the mode of the input
environment is Work, and $1 - $ InReceiveBit otherwise.

\item OutReceiveBit = OutSendBit if the mode of the output
environment is Ready, and $1 - $ OutSendBit otherwise.
\end{enumerate}
\end{enumerate}
\end{lem}

\begin{pf}
We check that $\rho$ is indeed a run of R2.  By the ordering lemma for
R1, the moves of every agent of R2 are linearly ordered.  It remains
to check only the coherence condition; the other conditions are
obvious.  Suppose that $Y$ is a finite initial segment of $N$ with a
maximal element $\mu$ and $X = Y - \{\mu\}$.  Using the facts that
$A(\mu)$ is enabled in $\sigma(X)$ and $\sigma(Y)$ is the result of executing
$A(\mu)$ in $\sigma(X)$, it is easy to check that $B(\mu)$ is enabled in
$\tau(X)$ and $\tau(Y)$ is the result of executing $B(\mu)$ at $\tau(X)$.
\qed \end{pf}

\begin{lem}
Conversely, every run of R2 is induced (in the sense of the preceding
lemma) by a unique run of R1.
\end{lem}

The proof is easy and we skip it.

\subsection{\Rea: The Official Row Evolving Algebra}
\label{officialRea}

After establishing that $p - g \neq N$ and before executing the
FrontCommunicate rule, buffer's front end goes to mode Ready.  This
corresponds to nothing in \RL\ which calls for merging the FrontWait
and FrontCommunicate rules.  On the other hand, \RL\ augments $p$ {\em
after\/} performing an act of communication.  There is no logical
necessity to delay the augmentation of $p$.  For aesthetic reasons we
merge the FrontWork rule with the other two rules of BuffFrontEnd.
Then we do a similar parallelization for BuffBackEnd.  Finally we
simplify the names BuffFrontEnd and BuffBackEnd to FrontEnd and
BackEnd respectively.  

A certain disaccord still remains because the environment is
implicit in \RL.  To remedy this, we remove the
environment modules, asserting that the functions InputDatum,
InSendBit, and OutReceiveBit which were updated by the environment
modules are now external functions.  The result is our official
ealgebra \Rea, shown in Figure~\ref{Rea}.

\begin{figure}[htbp]
\begin{tabbing}
\quad\=\qquad\=\qquad\=\kill
\rule{\textwidth}{1pt}\\

Module FrontEnd\\
\>\If $p - g \neq N$ \And InSendBit $\neq$ InReceiveBit \Then\\
\> \> Buffer($p \bmod N$) := InputDatum\\
\> \> InReceiveBit := 1 - InReceiveBit\\
\> \> $p$    := $p + 1$\\
\>\Endif\\
\rule{\textwidth}{1pt}\\

Module BackEnd\\
\>\If $p - g \neq 0$ \And OutSendBit $=$ OutReceiveBit \Then\\
\> \> OutputDatum := Buffer($g \bmod N$)\\
\> \> OutSendBit := 1 - OutSendBit\\
\> \> $g$        := $g + 1$\\
\>\Endif\\
\rule{\textwidth}{1pt}
\end{tabbing}
\caption{The program for \Rea.} \label{Rea}
\end{figure}

The initial states of \Rea\ satisfy the first condition on the initial
states of R1:\ The universe Integers and the arithmetical function
names mentioned in the program have their usual meanings; the universe
\ZN\ consists of integers modulo $N$ identified with the integers
$0,\ldots,N-1$; the universe \Ztwo\ is similar; $p = g = 0$; Buffer is
of type \ZN\ $\rightarrow$ Data; InputDatum and OutputDatum take values in
Data.

Additionally, the universe Agents contains two elements
to which Mod assigns different module names.  InSendBit, InReceiveBit,
OutSendBit, and OutReceiveBit are all equal to $0$.  InputDatum and
OutputDatum take values in Data.

The definition of regular runs of \Rea\ is slightly more complicated,
due to the presence of the external functions InputDatum, InSendBit,
and OutReceiveBit.  We require that the output sequence is at least as
long as the input sequence, InputDatum is of type Data, and InSendBit
and OutReceiveBit are both of type \Ztwo.

We skip the proof that \Rea\ is faithful to R2.

\subsection{\Cea: The Official Column Evolving Algebra}
\label{officialCea}

The evolving algebra \Cea\ is shown in figure \ref{Cea} below.  It
can be obtained from \CL\ in the same way that \Rea\ can be obtained 
from \RL; for brevity, we omit the intermediate stages.

\begin{figure}[htbp]
\begin{tabbing}
\quad\=\qquad\=\qquad\=\kill
\rule{\textwidth}{1pt}\\
Module Slot\\
\\
Rule Get\\
\> \If Mode(Me)=Get \And InputTurn(Me)\\
\> \> \> \And InSendBit $\neq$ InReceiveBit \Then\\
\> \> Buffer(Me) := InputDatum\\
\> \> InReceiveBit := 1 - InReceiveBit\\
\> \> $pp(\mbox{Me})$ := $1 - pp(\mbox{Me})$\\
\> \> Mode(Me) := Put\\
\> \Endif\\
\\
Rule Put\\
\> \If Mode(Me)=Put \And OutputTurn(Me)\\
\> \> \> \And OutSendBit = OutReceiveBit \Then\\
\> \> OutputDatum := Buffer(Me)\\
\> \> OutSendBit := 1 - OutSendBit\\
\> \> $gg(\mbox{Me})$ := $1 - gg(\mbox{Me})$\\
\> \> Mode(Me) := Get\\
\> \Endif\\
\\
InputTurn(x) abbreviates\\
\>[$x = 0$ \And $pp(0) = pp(N-1)$] \Or [$x \neq 0$ \And $pp(x) \neq
pp(x-1)$]\\
OutputTurn(x) abbreviates\\
\>[$x = 0$ \And $gg(0) = gg(N-1)$] \Or [$x \neq 0$ \And $gg(x) \neq
gg(x-1)$]\\
\rule{\textwidth}{1pt}
\end{tabbing}
\caption{The program for \Cea.} \label{Cea}
\end{figure}

\paragraph{Initial states}
The initial states of \Cea\ satisfy the following conditions.

\begin{enumerate}
\item 
The first condition for the initial states of R1 is satisfied except
we don't have functions $p$ and $g$ now.  Instead we have dynamic
functions $pp$ and $gg$ with domain \ZN\ and $pp(i) = gg(i) = 0$ for
all $i$ in \ZN.  

\item 
The universe Agents consists of the
elements of \ZN, which are mapped by Mod to the module name Slot.
Nullary functions Get and Put are distinct elements of the universe
Modes.  The dynamic function Mode is defined over Agents;
Mode$(x)$=Get for every $x$ in \ZN.  InputDatum and OutputDatum are
elements of Data.  Nullary functions InSendBit, InReceiveBit,
OutSendBit, OutReceiveBit are all equal to $0$.
\end{enumerate}

Regular runs are defined similarly to \Rea; we require that the output
sequence is at least as long as the input sequence, InputDatum is of
type Data, and InSendBit and OutReceiveBit take values in \Ztwo.

\section{Equivalence}
\label{equivpf}

We define a strong version of lock-step equivalence for ealgebras
which for brevity we call {\em lock-step equivalence}.  We
then prove that\ \Rea\ and\ \Cea\ are lock-step equivalent.  We start
with an even stronger version of lock-step equivalence which
we call {\em strict lock-step equivalence}.

For simplicity, we restrict attention to ealgebras with a fixed
superuniverse.  In other words, we suppose that all initial states
have the same superuniverse.  This assumption does not reduce 
generality because the superuniverse can be always chosen to be
sufficiently large.

\subsection{Strict Lock-Step Equivalence}

Let $\cal A$ and $\cal B$ be ealgebras with the same superuniverse and
suppose that $h$ is a one-to-one mapping from the states of $\cal A$
onto the states of $\cal B$ such that if $h(a) = b$ then $a$ and $b$
have identical interpretations of the function names common to $\cal
A$ and $\cal B$.  
Call a run $(M, A, \sigma)$ of $\cal A$
{\em strictly $h$-similar\/} to a partially ordered run $(N, B, \tau)$
of $\cal B$ if there is an isomorphism $\eta: M \rightarrow N$ such
that for every finite initial segment $X$ of $M$, $h(\sigma(X)) =
\tau(Y)$, where $Y = \{\eta(\mu): \mu \in X\}$.  Call $\cal A$ and
$\cal B$ {\em strictly $h$-similar\/} if every run of $\cal A$ is
strictly $h$-similar to a run of $\cal B$, and every run of $\cal B$
is $h^{-1}$-similar to a run of $\cal A$.  Finally call $\cal A$ and
$\cal B$ {\em strictly lock-step equivalent\/} if there exists an $h$
such that they are strictly $h$-similar.

Ideally we would like to prove that \Rea\ and \Cea\ are strictly
lock-step equivalent.  Unfortunately this is false, which is especially
easy to see if the universe Data is finite.  In this case, any run of
\Cea\ has only finitely many different states; this is not true for
\Rea\ because $p$ and $g$ may take arbitrarily large integer values.
One can rewrite either \Rea\ or \Cea\ to make them strictly lock-step
equivalent.  For example, \Cea\ can be modified to perform math on
$pp$ and $gg$ over Integers instead of \Ztwo.  We will not change
either ealgebra; instead we will slightly weaken the notion of strict
lock-step equivalence.

\subsection{Lock-Step Equivalence}

If an agent $\alpha$ of an ealgebra $\cal A$ is enabled at a state $a$, let
Result$(\alpha,a)$ be the result of firing $\alpha$ at $a$; otherwise
let Result$(\alpha,a) = a$.  

Say that an equivalence relation $\cong$ on the states of $\cal A$ {\em
respects\/} a function name $f$ of $\cal A$ if $f$ has the same
interpretation in equivalent states.  The equivalence classes of $a$
will be denoted $[a]$ and called the {\em configuration\/} of $a$.
Call $\cong$ a {\em congruence\/} if $a_1 \cong a_2 \rightarrow
\mbox{Result}(\alpha,a_1) \cong \mbox{Result}(\alpha,a_2)$ for any
states $a_1, a_2$ and any agent $\alpha$.

Let $\cal A$ and $\cal B$ be ealgebras with the same superuniverse and
congruences $\congA$ and $\congB$ respectively.  (We will drop the
subscripts on $\cong$ when no confusion arises.)  We suppose that
either congruence respects the function names common to $\cal A$ and
$\cal B$.  Further, let $h$ be a one-to-one mapping of
$\congA$-configurations onto $\congB$-configurations such that, for
every function name $f$ common to $\cal A$ and $\cal B$, if $h([a]) =
[b]$, then $f_a = f_b$.

Call a partially ordered run $(M, A, \sigma)$ of $\cal A$ {\em
$h$-similar\/} to a partially ordered run $(N, B, \tau)$ of $\cal B$
if there is an isomorphism $\eta: M \rightarrow N$ such that, for
every finite initial segment $X$ of $M$, $h([\sigma(X)]) = [\tau(Y)]$,
where $Y = \{\eta(\mu): \mu \in X\}$.  Call $\cal A$ and $\cal B$ {\em
$h$-similar\/} if every run of $\cal A$ is $h$-similar to a run of
$\cal B$, and every run of $\cal B$ is $h^{-1}$-similar to a run of
$\cal A$.  Call $\cal A$ and $\cal B$ {\em lock-step equivalent\/} (with
respect to $\congA$ and $\congB$) if there exists an $h$ such that
$\cal A$ and $\cal B$ are $h$-similar.

Note that strict lock-step equivalence is a special case of lock-step
equivalence, where $\congA$ and $\congB$ are both the identity
relation.

\bigskip

Assuming that \Rea\ and \Cea\ have the same superuniverse, we will show
that \Rea\ is lock-step equivalent to \Cea\ with respect to the
congruences defined below.  
\medskip

Remark.  The assumption that \Rea\ and \Cea\ have the same superuniverse
means essentially that the superuniverse of \Cea\ contains all integers
even though most of them are not needed.  It is possible to remove the
assumption.  This leads to slight modifications in the proof.  One
cannot require that a common function name $f$ has literally the same
interpretation in a state of \Rea\ and a state of \Cea.  Instead require
that the interpretations are essentially the same.  For example,
if $f$ is a predicate, require that the set of tuples where $f$ is
true is the same.      

\begin{defn}
For states $c,d$ of \Cea, $c \cong d$ if $c = d$.
\end{defn}

Since each configuration of \Cea\ has only one element, we identify a
state of \Cea\ with its configuration.  Let $e_a$ denote the value of
an expression $e$ at a state $a$.

\begin{defn}
For states $a,b$ of \Rea, $a \cong b$ if:
\begin{itemize}
\item $g_a = g_b\  \bmod 2N$
\item $(p-g)_a = (p-g)_b$
\item $f_a = f_b$ for all other function names $f$.
\end{itemize}
\end{defn}

Let $\div$ represent integer division: $i \div j = \lfloor i /
j \rfloor$.

\begin{lem} 
\label{lem0} 
If $a \congR b$ then we have the following modulo $2$:
\begin{itemize}
\item $p_a \div N$ = $p_b \div N$
\item $g_a \div N$ = $g_b \div N$
\end{itemize}
\end{lem}

\begin{pf}
We prove the desired property for $p$; the proof for $g$ is similar.

By the definition of $\congR$, we have the following modulo $2N$: 
$p_a = g_a + (p-g)_a = g_b + (p-g)_b = p_b$.  Thus, there are
non-negative integers $x_1, x_2, x_3, y$ such that $p_a = 2Nx_1 + Nx_2
+ x_3$, $p_b = 2Ny + Nx_2 + x_3$, $x_2 \leq 1$, and $x_3 < N$.  Hence
$p_a \div N = 2x_1 + x_2$ and $p_b \div N = 2y + x_2$, which are
equal modulo $2$.  \qed \end{pf}

\bigskip

We define a mapping $h$ from configurations of \Rea\ onto
configurations of \Cea.

\begin{defn}  If $a$ is a state of \Rea, then $h([a])$ is the state
$c$ of \Cea\ such that
\[
\begin{array}{rl}
 pp(i)_c = & \left\{ 
		\begin{array}{ll}
		p_a \div N \bmod 2 & \mbox{if } i \geq p_a \bmod N\\
		1 - (p_a \div N) \bmod 2 & \mbox{otherwise}
		\end{array} \right. \\
 gg(i)_c = & \left\{ 
		\begin{array}{ll}
		g_a \div N \bmod 2 & \mbox{if } i \geq g_a \bmod N\\
		1 - (g_a \div N) \bmod 2 & \mbox{otherwise}
		\end{array} \right. \\
\end{array}
\]
and for all common function names $f$, $f_c = f_a$.
\end{defn}

Thus, $h$ relates the counters $p,g$ used in \Rea\ and the counters $pp,gg$
used in \Cea.  (Notice that by Lemma \ref{lem0}, $h$ is well-defined.)  We
have not said anything about {\em Mode\/} because {\em Mode\/} is uniquely
defined by the rest of the state (see Lemma \ref{modelem} in section
\ref{proofs}) and is redundant.  

We now prove that \Rea\ and \Cea\ are $h$-similar.


\subsection{Properties of \Rea}
\label{proofs}

We say that $a$ is a state of a run $(M,A,\sigma)$ if $a=\sigma(X)$ for some
finite initial segment $X$ of $M$.

\begin{lem}
\label{lem1}
For any state $b$ of any run of \Rea, $0 \leq (p-g)_b \leq N$.
\end{lem}

\begin{pf} 
By induction. Initially, $p=g=0$.

Let $(M,A,\sigma)$ be a run of \Rea.  Let $X$ be a finite initial segment
of $M$ with maximal element $\mu$, such that $0 \leq p-g \leq N$ holds in $a
=\sigma(X - \{\mu\})$.  Let $b = \sigma(X)$.
\begin{itemize}
\item If $A(\mu)$ is the front end agent and is enabled in $a$, then
$0 \leq (p-g)_a < N$.  The front end agent increments $p$ but does not alter
$g$; thus, $0 < (p-g)_b \leq N$.
\item If $A(\mu)$ is the back end agent and is enabled in $a$, then
$0 < (p-g)_a \leq N$.  The back end agent increments $g$ but does not
alter $p$; thus, $0 \leq (p-g)_b < N$. \qed
\end{itemize}
\end{pf}


\begin{lem}
\label{lemk}
Fix a non-negative integer $k<N$.  For any run $(M, A, \sigma)$ of
\Rea, the k-slot moves of $M$ (that is, the moves of $M$ which involve
Buffer($k$)) are linearly ordered.
\end{lem}

\begin{pf} Similar to Lemma~\ref{orderingR1}.  \qed \end{pf}


\subsection{Properties of \Cea}

\begin{lem}
\label{lem2}
For any run of \Cea, there is a mapping In from states
of \Cea\ to \ZN\ such that if $In(c) = k$, then:
\begin{itemize}
\item {InputTurn}(Me) is true for agent $k$ and for no other agent.
\item For all $i < k$, $pp(i)_c = 1 - pp(k)_c$.
\item For all $k \leq i < N$, $pp(i)_c = pp(k)_c$. 
\end{itemize}
\end{lem}

\begin{pf} 
By induction.  Initially, agent $0$ (and no other) satisfies
{\em InputTurn(Me)\/} and $pp(i)=0$ holds for every
agent $i$.  Thus, if $c$ is an initial state, $In(c) = 0$.

Let $(M,A,\sigma)$ be a run of \Cea.  Let $Y$ be a finite initial segment of
$M$ with maximal element $\mu$, such that the requirements hold in $c =
\sigma(Y - \{\mu\})$.  Let $d = \sigma(Y)$.   

If $A(\mu)$ executes rule Put, $pp$ is not modified and $In(d) = In(c)$.
Otherwise, if rule Get is enabled for $A(\mu)$, executing rule Get increments
$pp$; the desired $In(d) = In(c) + 1 \bmod N$.  This is obvious if $In(c) <
N-1$.  If $In(c) = N-1$, then all values of $pp$ are equal in $d$ and
$In(d)=0$ satisfies the requirements.
\qed \end{pf}


\begin{lem}
\label{lem3}
For any run of \Cea, there is a mapping Out from states
of \Cea\ to \ZN\ such that if $Out(c) = k$, then:
\begin{itemize}
\item {OutputTurn}(Me) is true for agent $k$ and no other agent.
\item For all $i < k$, $gg(i)_c = 1 - gg(k)_c$.
\item For all $k \leq i < N$, $gg(i)_c = gg(k)_c$. 
\end{itemize}
\end{lem}

\begin{pf} Parallel to that of the last lemma.  \qed \end{pf}


It is easy to see that every move $\mu$ of \Cea\ involves an execution
of rule Get or rule Put but not both.  (More precisely, consider
finite initial segments $Y$ of moves where $\mu$ is a maximal element
of $Y$.  Any such $Y$ is obtained from $Y - \{\mu\}$ either by
executing Get in state $\sigma(Y-\{\mu\})$, or executing Put in
state $\sigma(Y-\{\mu\})$.)  In the first case, call $\mu$ a Get move.
In the second case, call $\mu$ a Put move.

\begin{lem}
\label{lem4}
In any run $(M, A, \sigma)$ of \Cea, all Get moves are linearly ordered
and all Put moves are linearly ordered.
\end{lem}

\begin{pf}
We prove the claim for rule Get; the proof for rule Put is similar.
By contradiction, suppose that are two incomparable Get moves $\mu$
and $\nu$.  By the coherence condition for runs, both rules are enabled
in state $X = \{\pi: \pi < \mu \vee \pi < \nu\}$.  By Lemma
\ref{lem2}, A($\mu$) = A($\nu$).  But all moves of the same agent are
ordered; this gives the desired contradiction.  \qed \end{pf}


\begin{lem}
\label{lem5}
\label{modelem}
In any state $d$ of any run of \Cea, for any agent k,
\[
Mode(k)_d = \left\{ \begin{array}{ll}
		Get & \mbox{if } pp(k)_d = gg(k)_d\\
		Put & \mbox{if } pp(k)_d = 1 - gg(k)_d
	   \end{array} \right.
\]
\end{lem}

\begin{pf}
We fix a $k$ and do induction over runs.  Initially, $Mode(k) =
Get$ and $pp(k) = gg(k) = 0$ for every agent $k$.

Let $Y$ be a finite initial segment of a run with maximal element
$\mu$ such that (by the induction hypothesis) the required condition
holds in $c = \sigma(Y - \{\mu\})$.  Let $d = \sigma(Y)$.

If $A(\mu) \neq k$, none of $Mode(k)$, $pp(k)$, and $gg(k)$ are
affected by executing $A(\mu)$ in $c$, so the condition holds in $d$.
If $A(\mu)=k$, we have two cases.

\begin{itemize}
\item
If agent $k$ executes rule Get in state $c$, we must have
$Mode(k)_c = Get$ (from rule Get) and $pp(k)_c = gg(k)_c$ (by the
induction hypothesis).
Firing rule Get yields $Mode(k)_d = Put$ and $pp(k)_d = 1 - pp(k)_c =
1 - gg(k)_d$.

\item
If agent $k$ executes rule Put in state $c$, we must have
$Mode(k)_c = Put$ (from rule Put) and $pp(k)_c = 1 - gg(k)_c$ (by the
induction hypothesis).
Firing rule Get yields $Mode(k)_d = Get$ and $gg(k)_d = 1 - gg(k)_c =
pp(k)_d$.  \qed 
\end{itemize}
\end{pf}

Remark.  This lemma shows that function {\em Mode\/} is indeed redundant.


\subsection{Proof of Equivalence}

\begin{lem}
\label{leminout}
If $h([a]) = c$, then $In(c) = p_a \bmod N$ and $Out(c) = g_a \bmod N$.
\end{lem}

\begin{pf}
Recall that {\em In(c)\/} is the agent $k$ for which {\em {InputTurn}(k)}$_c$
holds.  Lemma \ref{lem2} asserts that $pp(i)_c$ has one value for $i < k$ and
another for $i \geq k$.  By the definition of $h$, this ``switch-point'' in
$pp$ occurs at $p_a \bmod N$.  The proof for $Out(c)$ is similar.
\qed \end{pf}


\begin{lem}
\label{lem11}
Module FrontEnd is enabled in state $a$ of \Rea\ iff rule Get is
enabled in state $c = h([a])$ of \Cea\ for agent $In(c)$.
\end{lem}

\begin{pf}
Let $k = In(c)$, so that {\em {InputTurn}(k)$_c$} holds.  Both FrontEnd and
Get have {\em InSendBit $\neq$ InReceiveBit\/} in their
guards.  It thus suffices to show that $(p-g)_a \neq N$ iff
$Mode(k)_c$ = Get.  By Lemma \ref{lem5}, it suffices to show that 
$(p-g)_a \neq N$ iff $pp(k)_c = gg(k)_c$.

Suppose $(p-g) \neq N$.  There exist non-negative integers $x_1, x_2, x_3,
x_4$ such that $p_a = x_1N + x_3$, $g_a = x_2N + x_4$, and $x_3, x_4 < N$.
(Note that by Lemma \ref{leminout}, $k = p_a \bmod N = x_3$.)

By Lemma \ref{lem1}, $0 \leq (p-g)_a < N$.   There are two cases.

\begin{itemize}
\item $x_1 = x_2$ and $x_3 \geq x_4$.  
By definition of $h$, we have that, modulo 2,
$pp(x_3)_c = p_a \div N = x_1$ and 
for all $i \geq g_a \bmod N = x_4$, $gg(i)_c = g_a \div N = x_2$.
Since $x_3 \geq x_4$, we have that, modulo 2,
$gg(x_3)_c = x_2 = x_1 = pp(x_3)_c$, as desired.

\item $x_1 = (x_2 + 1)$ and $x_3 < x_4$.  By definition of $h$, 
we have that, modulo 2, $pp(x_3)_c = p_a \div N = x_1$
and for all $i < g_a \bmod N = x_4$,
$gg(i)_c$ = 1 - $g_a \div N = x_2 + 1$.
Since $x_3 < x_4$, we have that, modulo 2,
$gg(x_3)_c = x_2 + 1 = x_1 = pp(x_3)_c$, as desired.
\end{itemize}

On the other hand, suppose $(p-g)_a = N$.  Then $p_a \div N$ and $g_a\ 
\div N$ differ by 1.  By definition of $h$, $pp(i)_c = 1-gg(i)_c$ for
all $i$, including $k$.
\qed \end{pf}


\begin{lem}
\label{lem12}
Module BackEnd is enabled in state $a$ iff rule Put is
enabled in state $c = h([a])$ for agent $Out(c)$.
\end{lem}

\begin{pf} Similar to that of the last lemma. 
\qed \end{pf}


\begin{lem}
\label{lem13}
Suppose that module FrontEnd is enabled in a state $a$ of \Rea\ for
the front end agent $I$ and rule Get is enabled in a state $c = h([a])$ of
\Cea\ for agent $In(c)$.  Let $b = Result(I,a)$ and $d = Result(In(c),
c)$.  Then $d = h([b])$.
\end{lem}

\begin{pf}
We check that $h([b]) = d$.

\begin{itemize}
\item Both agents execute
{\em InReceiveBit := 1 -- InReceiveBit}.
\item The front end agent executes {\em Buffer($p$ mod N) := InputDatum}.
Agent $In(c)$ executes {\em Buffer(In(c)) := InputDatum}.
By Lemma \ref{leminout}, {\em In(c) = $p_a \bmod N$}, so
these updates are identical.
\item The front end agent executes $p := p + 1$.  Agent $In(c)$ executes
$pp(In(c)) := 1 - pp(In(c))$.  The definition of $h$ and the fact
that $pp(i)_c = pp(i)_{h([a])}$ for all $i \in {\cal Z}_N$
imply that $pp(i)_d = pp(i)_{h([b])}$.
\item Agent $In(c)$ executes {\em Mode(In(c)) := Put}.  By Lemma \ref{lem5},
this update is redundant and need not have a corresponding update
by the front end agent.  \qed
\end{itemize}
\end{pf}


\begin{lem}
\label{lem14}
Suppose that module BackEnd is enabled in a state $a$ of \Rea\ for the
back end agent $O$ and rule Put is enabled in a state $c = h([a])$ of \Cea\
for agent $Out(c)$.  Let $b = Result(O,a)$ and $d = Result(Out(c),
c)$.  Then $d = h([c])$.
\end{lem}

\begin{pf} Parallel to that of the last theorem.  \qed \end{pf}


\begin{thm} \label{isequiv}
\Rea\ is lock-step equivalent to \Cea.
\end{thm}

\begin{pf} 
Let $\Lambda(\mu) = \Lambda_{\cal R}(\mu)$ and
$\Lambda'(\mu) = \Lambda_{\cal C}(\mu)$.

We begin by showing that any run $(M, A, \sigma)$ of \Rea\ is $h$-similar
to a run of \Cea, using the definition of $h$ given earlier.
Construct a run $(M, A', \sigma')$ of \Cea, where $\sigma'(X) = h([\sigma(X)])$
and $A'$ is defined as follows.  Let $\mu$ be a move of $M$, $a =
\Lambda(\mu)$, and $c = h([\Lambda(\mu)])$.  Then $A'(\mu) = In(c)$ if
$A(\mu)$ is the front end agent, and $A'(\mu) = Out(c)$ if $A(\mu)$ is
the back end agent.

We check that $(M, A', \sigma')$ satisfies the four requirements
for a run of \Cea\ stated in Section \ref{runs}.
\begin{enumerate}
\item Trivial, since $(M, A, \sigma)$ is a run.

\item By Lemma \ref{lemk}, it suffices to show that for any
$\mu$, if $A'(\mu) = k$, then $A(\mu)$ is a $k$-slot move.  By the
construction above and Lemma \ref{leminout}, we have modulo N that $k
= In(c) = p_a$ if $A(\mu)$ is the front end agent and $k
= Out(c) = g_a$ if $A(\mu)$ is the back end agent.  In
either case, $\mu$ is a $k$-slot move.

\item Since $\sigma' = h \circ \sigma$, $\sigma'$  maps
finite initial segments of $M$ to states of \Cea.

\item {\em Coherence}.  
Let $Y$ be a finite initial segment of $M$ with a maximal element $\mu$,
and $X = Y - \{\mu\}$.  Thus {\em Result(A($\mu$),$\sigma$(X)) = $\sigma$(Y)}.
By Lemma \ref{lem11} or \ref{lem12}, $A'(\mu)$ is enabled
in $\sigma'(X)$.  By Lemma \ref{lem13} or \ref{lem14},
{\em Result($A'(\mu),\sigma'(X)$) = $\sigma'(Y)$}.
\end{enumerate}

Continuing, we must also show that for any run $(M, A', \sigma')$
of \Cea, there is a run $(M, A, \sigma)$ of \Rea\
which is $h$-similar to it.  

We define $A$ as follows.  Consider the action of agent
$A'(\mu)$ at state $\Lambda'(\mu)$.  If $A'(\mu)$ executes
rule Get, set $A(\mu)$ to be the front end agent.  If $A'(\mu)$ executes
rule Put, set $A(\mu)$ to be the back end agent.  

We check that the moves of the front end agent are linearly ordered.
By Lemma \ref{lem4}, it suffices to show that if $A(\mu)$ is the front
end agent, then $A'(\mu)$ executes Get in state $\Lambda'(\mu)$ ---
which is true by construction of $A$.  A similar argument shows
that the moves of the back end agent are linearly ordered.

We define $\sigma$ inductively over finite initial segments of $M$.
$\sigma(\emptyset)$ is the unique initial state in
$h^{-1}(\sigma'(\emptyset))$.

Let $Y$ be a finite initial segment with a maximal element $\mu$ such
that $\sigma$ is defined at $X = Y - \{\mu\}$.  Choose $\sigma(Y)$
from $h^{-1}(\sigma'(Y))$ such that $\sigma(Y)^- =
Result(A(\mu),\sigma(X))$.  Is it possible to select such a
$\sigma(Y)$?  Yes.  By Lemma \ref{lem11} or \ref{lem12}, $A(\mu)$ is
enabled in $\sigma(X)$ iff $A'(\mu)$ is enabled in $\sigma'(X)$. By
Lemma \ref{lem13} or \ref{lem14}, {\em Result($A(\mu),\sigma(X)$)
$\in$ $h^{-1}$(Result($A'(\mu),\sigma'(\mu)$))}.
It is easy to check that $(M,A,\sigma)$ is a run of \Rea\ which is
$h$-similar to $(M,A',\sigma')$.  \qed 
\end{pf}

\section{Inequivalence}
\label{inequiv}

We have proven that our formalizations \Rea\ and \Cea\ of 
\RL\ and \CL\ are lock-step equivalent.  Nevertheless, \RL\ and
\CL\ are inequivalent in various other ways.  In the following
discussion we exhibit some of these inequivalences.  The discussion
is informal, but it is not difficult to prove these inequivalences
using appropriate formalizations of \RL\ and \CL.  Let
${\cal R} = {\cal R}_{\mathrm{pcsp}}$ and 
${\cal C} = {\cal C}_{\mathrm{pcsp}}$.

\paragraph*{Magnitude of Values.}  ${\cal R}$ uses unrestricted integers 
as its counters; in contrast, ${\cal C}$ uses only single bits for the
same purpose.  We have already used this phenomenon to show that \Rea\
and \Cea\ are not strictly lock-step equivalent.  One can put the same
argument in a more practical way.  Imagine that the universe Data is
finite and small, and that a computer with limited memory is used to
execute ${\cal R}$ and ${\cal C}$.  ${\cal R}$'s counters may
eventually exceed the memory capacity of the computer.  ${\cal C}$
would have no such problem.

\paragraph*{Types of Sharing.}  ${\cal R}$ shares access to the
buffer between both processes; in contrast, each process in ${\cal C}$ has
exclusive access to its portion of the buffer.  Conversely, processes
in ${\cal C}$ share access to both the input and output channels, while each
process in ${\cal R}$ has exclusive access to one channel.  Imagine an
architecture in which processes pay in one way or another for
acquiring a channel.  ${\cal C}$ would be more expensive to use on such a
system.

\paragraph*{Degree of Sharing.}  
How many internal locations used by each algorithm must be shared
between processes?  ${\cal R}$ shares access to $N+2$ locations: the
$N$ locations of the buffer and $2$ counter variables.  ${\cal C}$
shares access to $2N$ locations: the $2N$ counter variables.  Sharing
locations may not be without cost; some provision must be made for
handling conflicts ({\em e.g.\/} read/write conflicts) at a given
location.  Imagine that a user must pay for each shared location (but
not for private variables, regardless of size).  In such a scenario,
${\cal C}$ would be more expensive than ${\cal R}$ to run.

\bigskip

These contrasts can be made a little more dramatic.  For example, one
could construct another version of the ring buffer algorithm which
uses $2N$ processes, each of which is 
responsible for an input or output action (but not both) to a
particular buffer position.  All of the locations it uses
will be shared.  It is lock-step equivalent to ${\cal R}$ and ${\cal C}$; yet, few
people would choose to use this version 
because it exacerbates the disadvantages of ${\cal C}$.  Alternatively, one could
write a single processor (sequential) algorithm which is equivalent in
a different sense to ${\cal R}$ and ${\cal C}$; it would produce the same output as
${\cal R}$ and ${\cal C}$ when given the same input but would
have the disadvantage of not allowing all
orderings of actions possible for ${\cal R}$ and ${\cal C}$.

\paragraph{Acknowledgements.}
We thank S\symbol{'034}ren B\symbol{'034}gh Lassen, Peter Mosses,
and the anonymous referees for their comments.


\end{document}